\documentclass[prc,superscriptaddress,noshowpacs,unsortedaddress,twocolumn,showpacs,preprintnumbers,amsmath,amssymb]{revtex4-1}

\usepackage[dvipdfmx]{graphicx}
\usepackage{amsmath,amssymb,times}
\usepackage{color}
\usepackage{ulem}
\usepackage{here}

\newcommand{\beq}{\begin{equation}}
\newcommand{\eeq}{\end{equation}}
\newcommand{\bea}{\begin{eqnarray}}
\newcommand{\eea}{\end{eqnarray}}

\newcommand{\bfi}[1]{\mbox{\boldmath $#1$}}

\newcommand{\vK}{{\bfi K}}

\newcommand{\vs}{{\bfi s}}

\newcommand{\vrr}{{\bfi r}}
\newcommand{\vR}{{\bfi R}}

\def\a{\alpha}

\begin{document}
\title{Matter radii and skins of $^{6,8}$He  from  reaction cross section of proton+$^{6,8}$He scattering
 \\
based on the Love-Franey $t$-matrix model}

\author{Tomotsugu~Wakasa}
\affiliation{Department of Physics, Kyushu University, Fukuoka 819-0395, Japan}

\author{Maya~Takechi}
\affiliation{Niigata University, Niigata 950-2181, Japan}

\author{Shingo~Tagami}
\affiliation{Department of Physics, Kyushu University, Fukuoka 819-0395, Japan}

\author{Masanobu Yahiro}
\email[]{orion093g@gmail.com}
\affiliation{Department of Physics, Kyushu University, Fukuoka 819-0395, Japan}             

\date{\today}

\begin{abstract}
\noindent 
{\bf Background:}
For $^{4,6,8}$He,  Tanihata {\it et al.} determined matter radii 
$r_{m}(\sigma_{\rm I})=1.57(4), 2.48(3), 2.52(3)$~fm 
from interaction cross sections $\sigma_{\rm I}$ for $^{4,6,8}$He scattering on Be, C Al targets 
at 790~MeV/nucleon. 
Lu {\it et al.} measured the atomic isotope shifts (AIS) for  $^{4,6,8}$He and 
determined proton radii $r_{p}({\rm AIS})$ for $^{4,6,8}$He. 
As for p+$^{4,6,8}$He scattering, reaction cross sections $\sigma_{\rm R}({\rm exp})$ 
are available at 700~MeV with high accuracy. 
\\
{\bf Aim:} 
Our aim is to determine matter radii  $r_{m}$ and skins $r_{\rm skin}$ for 
$^{6,8}$He  from the $\sigma_{\rm R}({\rm exp})$ and the $r_{p}({\rm AIS})$.  
\\
{\bf Method:} 
Our model is the Love-Franey $t$-matrix folding model, since the model is better than the optical limit of 
Glauber model.  
\\
{\bf Results:} 
Our results for $^{6,8}$He are $r_{m}({\rm exp})=2.48(3), 2.53(2)$~fm and 
$r_{\rm skin}=$0.78(3), 0.82(2)~fm. 
\\
{\bf Conclusion:} 
For $^{6,8}$He, our results $r_{m}(\sigma_{\rm R})$ 
agree with  those of Tanihata {\it et al.}. 
For $^{8}$He, the  distance between $^{4}$He and the center of mass of valence four neutrons is 2.367~fm.
\end{abstract}

\maketitle


\section{Introduction and conclusion}
\label{Introduction}

{\it Background:}
The matter radius $r_{m}$, the neutron skin $r_{\rm skin}$ and halo structure 
are important properties of nuclei. 
When a nucleus has one or more loosely-bound nucleons surrounding a tightly bound core, 
it is considered that the nucleus has a halo structure. Eventually, we may consider that   
$^{6,8}$He have the halo structure.  

 Lu {\it et al.} measured the atomic isotope shifts (AIS) along $^{4,6,8}$He by performing laser spectroscopy 
on individual trapped atoms and determined proton radii as $r_{p}({\rm AIS})=1.462(6), 1.934(9), 1.881(17)$~fm for 
$^{4,6,8}$He~\cite{Lu:2013ena}. 

For He isotopes, meanwhile, Tanihata {\it et al.} determined $r_{m}$ from interaction cross sections $\sigma_{\rm I}$ 
for $^{4,6,8}$He scattering of Be, C Al targets at 790~MeV/nucleon~\cite{Tanihata:1988ub};  
their results are $r_{m}(\sigma_{\rm I})=1.57(4), 2.48(3), 2.52(3)$~fm for $^{4,6,8}$He 
in which the the harmonic-oscillator distribution is assumed for the densities for $^{4,6,8}$He. 
They used the optical limit of Glauber model~\cite{Glauber,Yahiro-Glauber}. 
The folding model is better than the optical limit of  the Glauber model, when the incident energy is 
smaller than nucleon mass.

As for p+$^{4,6,8}$He scattering, the  data on reaction cross section $\sigma_{\rm R}$ are available 
at 700~MeV~\cite{Neumaier:2002eay} with high accuracy of 1.7\%.
In Ref.~~\cite{Neumaier:2002eay}, absolute differential cross sections for elastic $^{4,6,8}$He 
 small-angle scattering were measured in inverse kinematics.

{\it Aim:} 
Our aim is to determine  matter radius $r_{m}$ and and skins $r_{\rm skins}$ for $^{6,8}$He  
from the data $\sigma_{\rm R}({\rm exp})$~\cite{Neumaier:2002eay} for p+$^{6,8}$He scattering at 700~MeV and 
the $r_{p}({\rm AIS})$, since the $\sigma_{\rm R}({\rm exp})$ have small errors of 1.7\%.  

{\it Method:} 
Our model is  the Love-Franey (LF) $t$-matrix folding model. 
We have already shown that the  folding model based on LF $t$-matrix~\cite{LF} is good 
for  $^{4,6,8}$He+$^{12}$C at 790~MeV per nucleon~\cite{Tagami:2020ajd} that is to be published in 
Results in Physics.  

{\it Results:} 
Our results for $^{6,8}$He are $r_{m}({\rm exp})=2.48(3), 2.53(2)$~fm and $r_{\rm skin}=0.78(3), 0.82(2)$~fm. 

{\it Conclusion:} 
For $^{6,8}$He, our results agree with  those of Tanihata {\it et al.} based on $\sigma_{\rm I}$.
For $^{8}$He, the  distance $d_{\a-4n}$ between $^{4}$He and the center of mass (cm) of valence four neutrons is 2.367~fm.

\section{Model}
\label{Sec-Framework}

We use the folding model  based on Lovey-dovey (LF) $t$-matrix~\cite{LF}.

We show the formulation on  the LF folding  $t$-matrix model below. 
For proton-nucleus scattering, the potential $U(\vR)$ 
 between a projectile (P)  and a target (${\rm T}$) has the direct and exchange parts,
$U^{\rm DR}$ and $U^{\rm EX}$, as
\begin{subequations}
\begin{eqnarray}
U^{\rm DR}(\vR) & = & 
\sum_{\mu,\nu}\int             \rho^{\nu}_{\rm T}(\vrr_{\rm T})
            t^{\rm DR}_{\mu\nu}(s;\rho_{\mu\nu})  d
	    \vrr_{\rm T}\ ,\label{eq:UD} \\
U^{\rm EX}(\vR) & = & 
\sum_{\mu,\nu}
\int \rho^{\nu}_{\rm T}(\vrr_{\rm T},\vrr_{\rm T}+\vs) \nonumber \\
                &   &
\times t^{\rm EX}_{\mu\nu}(s;\rho_{\mu\nu}) \exp{[-i\vK(\vR) \cdot \vs/M]}
             d \vrr_{\rm T}\,~~
             \label{eq:UEX}
\end{eqnarray}
\end{subequations}
where $\vR$ is the relative coordinate between P  and T,
$\vs=-\vrr_{\rm T}+\vR$, and $\vrr_{\rm T}$ is
the coordinate of the interacting nucleon from T.
 Each of $\mu$ and $\nu$ denotes the $z$-component of isospin. 
The nonlocal $U^{\rm EX}$ has been localized in Eq.~\eqref{eq:UEX}
with the local semi-classical approximation~\cite{Brieva-Rook}
where \vK(\vR) is the local momentum between P and T, 
and $M= A/(1 +A)$ for the target mass number $A$;
see Ref.~\cite{Minomo:2009ds} for the validity of the localization.

The direct and exchange parts, $t^{\rm DR}_{\mu\nu}$ and 
$t^{\rm EX}_{\mu\nu}$, of the $t$ matrix are described by
\begin{align}
t_{\mu\nu}^{\rm DR}(s) 
&=
\displaystyle{\frac{1}{4} \sum_S} \hat{S}^2 t_{\mu\nu}^{S1}
 (s) \hspace*{0.1cm}  \hspace*{0.1cm} 
 {\rm for} \hspace*{0.1cm} \mu+\nu = \pm 1,
 \\
t_{\mu\nu}^{\rm DR}(s) 
&=
\displaystyle{\frac{1}{8} \sum_{S,T}} 
\hat{S}^2 t_{\mu\nu}^{ST}(s) 
\hspace*{0.1cm}  \hspace*{0.1cm} 
{\rm for} \hspace*{0.1cm} \mu+\nu = 0,
\\
t_{\mu\nu}^{\rm EX}(s) 
&=
\displaystyle{\frac{1}{4} \sum_S} (-1)^{S+1} 
\hat{S}^2 t_{\mu\nu}^{S1} (s) 
\hspace*{0.1cm}  \hspace*{0.1cm} 
{\rm for} \hspace*{0.1cm} \mu+\nu = \pm 1, 
\\
t_{\mu\nu}^{\rm EX}(s) 
&=
\displaystyle{\frac{1}{8} \sum_{S,T}} (-1)^{S+T} 
\hat{S}^2 t_{\mu\nu}^{ST}(s) 
\hspace*{0.1cm}  \hspace*{0.1cm}
{\rm for} \hspace*{0.1cm} \mu+\nu = 0 
,
\end{align}
where $\hat{S} = {\sqrt {2S+1}}$ and $t_{\mu\nu}^{ST}$ are 
the spin-isospin components of the $t$-matrix interaction.
We apply the LF $t$-matrix  folding model for  p+$^{4,6,8}$He scattering   
at $E_{\rm in}=700$~MeV.

As proton and neutron densities, $\rho^{\nu=-1/2}_{\rm T}$ and $\rho^{\nu=1/2}_{\rm T}$, 
 we use the densities calculated with D1S-Gogny HFB (D1S-GHFB)~\cite{Tagami:2019svt}. 
 As a way of taking the center-of-mass correction to the densities, 
we use the method of Ref.~\cite{Sumi:2012fr}.  
We scale  D1S-GHFB proton and  neutron densities, as mentioned below. 

We consider proton and neutron densities calculated with D1S-GHFB as 
the original density $\rho(\vrr)$. 
The scaled density $\rho_{\rm scaling}(\vrr)$ is determined  
from the original density $\rho(\vrr)$ as
\bea
\rho_{\rm scaling}(\vrr) \equiv \frac{1}{\a^3}\rho(\vrr/\a), ~\vrr_{\rm scaling} \equiv \vrr/\a
\label{eq:scaling}
\eea
with a scaling factor
\bea
\a=\sqrt{ \frac{\langle \vrr^2 \rangle_{\rm scaling}}{\langle \vrr^2 \rangle}} .
\eea
In Eq.~\eqref {eq:scaling}, we have replaced $\vrr$ by $\vrr/\a$ in the original density. 
Eventually, $\vrr$ dependence 
of   $\rho_{\rm scaling}(\vrr)$ is different from that of  $\rho(\vrr)$. 
We have multiplied the original density by $\a^{-3}$ 
 in order to normalize the scaled density. 
The symbol means $\sqrt{\langle \vrr^2 \rangle_{\rm scaling}}$ is the root-mean-square
radius of  $\rho_{\rm scaling}(\vrr)$. 

For later convenience, we refer to the proton (neutron) radius of the scaled proton (neutron) density $\rho^{\rm p}_{\rm scaling}(\vrr)$ 
($\rho^{\rm n}_{\rm scaling}(\vrr)$) as $r_{\rm p}({\rm scaling})$ ($r_{\rm n}({\rm scaling})$).

\section{Results}
\label{Results}

For $^{6,8}$He, we first deduce neutron radius $r_{n}(\sigma_{\rm I})=2.71,  2.70$~fm  from 
the $r_{m}(\sigma_{\rm I})=2.48,2.52$~fm and the $r_{p}({\rm AIS})=1.934, 1.881$~fm. 
For  $^{4}$He,  we assume $r_{n}({\rm AIS})=r_{p}({\rm AIS})$, i.e.,   
$r_{m}({\rm AIS})=r_{n}({\rm AIS})=r_{p}({\rm AIS})$. 
For $^{6,8}$He, the $r_{n}(\sigma_{\rm I})$ and the $r_{p}({\rm AIS})$ yields $r_{m}({\rm exp})=2.48(3), 2.53(3)$~fm. 

\begin{figure}[H]
\begin{center}
 \includegraphics[width=0.5\textwidth,clip]{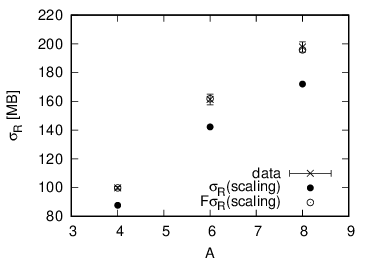}
 \caption{ 
Reaction cross sections $\sigma_{\rm R}$  for $p$+$^{4,6,8}$He scattering at 700~MeV. 
 Closed circles denote results  $\sigma_{\rm R}({\rm scaling})$ of the  scaled densities based on 
 $r_{p}({\rm scaling})=r_{p}({\rm AIS})$ and 
$r_{n}({\rm scaling})=r_{n}({\rm AIS})$ for $^{4}$He and 
$r_{p}({\rm scaling})=r_{p}({\rm AIS})$ and 
$r_{n}({\rm scaling})=r_{n}(\sigma_{\rm I})$ for $^{6,8}$He.  
Open circles correspond  to $F \sigma_{\rm R}({\rm scaling})$. 
 The data (crosses) are taken from Ref.~\cite{Neumaier:2002eay}.
   }
 \label{Fig-Rx sec-p+He}
\end{center}
\end{figure}

For $^{4,6,8}$He,  we scale 
proton and neutron  D1S-GHFB densities so as to satisfy $r_{p}({\rm scaling})=r_{p}({\rm AIS})$ and 
$r_{n}({\rm scaling})=r_{n}({\rm AIS})$ for $^{4}$He and 
$r_{p}({\rm scaling})=r_{p}({\rm AIS})$ and 
$r_{n}({\rm scaling})=r_{n}(\sigma_{\rm I})$ for $^{6,8}$He.  
For $^{4,6,8}$He, the reaction cross section $\sigma_{\rm R}({\rm scaling})$ calculated 
with the scaled densities undershoot the $\sigma_{\rm R}({\rm exp})$ by 12\%, 
as shown in Fig.~\ref{Fig-Rx sec-p+He}.

For $^{4}$He, we introduce the fine-tuning factor $F$ as 
$F=\sigma_{\rm R}({\rm exp})/\sigma_{\rm R}({\rm scaling})=1.1385$. 
This fine-tuning is necessary  for light projectiles and targets~\cite{Tagami:2020ajd}. 
The $F\sigma_{\rm R}({\rm scaling})$ reproduce $\sigma_{\rm R}({\rm exp})$ 
for $^{4,6,8}$He, as shown in Fig.~\ref{Fig-Rx sec-p+He} for $\sigma_{\rm R}({\rm exp})$ of p+$^{4,6,8}$He 
at 700~MeV. 
For $^{6,8}$He, we scale the proton and neutron D1S-GHFB densities so as to 
$F\sigma_{\rm R}({\rm scaling})=\sigma_{\rm R}({\rm exp})$ and 
$r_{p}({\rm scaling})=r_{p}({\rm AIS})$.
Therefore, our results based on the scaling method are $r_{m}({\rm exp})=2.48(3), 2.53(2)$~fm and 
$r_{\rm skin}=$0.78(3), 0.82(2)~fm for $^{6,8}$He.

The proton radius of $^{6}$He comes from the proton radius of $^{4}$He and the  distance 
$d_{\a-2n}$ between $^{4}$He and the cm of valence two neutron; 
namely, 
\bea
r_{p}({\rm AIS},{^6}{\rm He})^2=r_{p}({\rm AIS},{^4}{\rm He})^2+\Big(\frac{2}{6}\Big)^2  
r_{\a-2n}^2 
\label{He6-distance}
\eea
The latter term represents the recoil effect of the cm.
The resulting $r_{\a-2n}$ is 3.798~fm, while the $^{4}$He+n+n model of Ref. \cite{Hiyama96} 
yields 3.79~fm.

For  $^{8}$He, the relation becomes 
\bea
r_{p}({\rm AIS},{^8}{\rm He})^2=r_{p}({\rm AIS},{^4}{\rm He})^2+\Big(\frac{4}{8}\Big)^2 
r_{\a-4n}^2 
\eea
The resulting $r_{\a-4n}$ is 2.367~fm.

\noindent
\appendix

\noindent
\begin{acknowledgments}
We would like to thank Dr. Toyokawa for his contribution. 
\end{acknowledgments}



\bibliographystyle{pasty}

\end{document}